\documentstyle [12pt] {article}
\begin{document}
\begin{flushright}
\begin{tabular}{r}
\hbox{\bf LPC 95/10}\cr
\hbox{hep-ph/9503421}\cr
\hbox{March 95}\cr
\end{tabular}
\end{flushright}
\vspace{2.0cm}
\begin{center}
{\bf STUDY OF $J/\Psi$ THREE-BODY DECAYS }\\[15pt]
{\bf INVOLVING BARYONS}\\[40pt]
{\bf El-Hassan KADA}\\[10pt]
{\bf D\'{e}partement de Physique - Facult\'{e} des Sciences}\\[10pt]
{\bf Oujda, Maroc}\\[30pt]
{\bf Joseph PARISI}\\[10pt]
{\bf Laboratoire de Physique Corpusculaire, Coll\`{e}ge de France}
\\[10pt]
{\bf IN2P3 - CNRS}\\[10pt]
{\bf 11 Place M. Berthelot, F-75231 Paris Cedex 05, France}
\end{center}
\vspace{1.0cm}

\noindent \underline{Abstract} : The $J/\Psi$ decay into a baryon
pair and a
pseudoscalar meson is computed, for some channels, in lowest order in
perturbative QCD, modeling the baryon with a quark-diquark system. We
use a set of parameters that has been proposed by some
authors in order to fit the proton magnetic form factor $G^{p}_{M}$,
the
angular distribution of protons in the process
$\gamma \gamma \rightarrow p \bar{p}$ and the width of
$\eta_{c} \rightarrow \gamma \gamma$.\\[30pt]
\vspace{1.0cm}

\newpage
\section{Introduction}
Several experimental data of inclusive processes involving nucleons,
as well as
theoretical indications, strongly suggest that, at intermediate
momentum
transfer $Q^{2}$, diquarks induced by strong two-quark correlations
inside baryons can behave like quasi-elementary constituents. The
same scheme
has been used in the description of several exclusive processes {\bf
[1]}.
\\[5pt]

A similar, although simplified, quark-diquark model has been applied
to the
description of the decay of the $J/\Psi$ into baryon pairs {\bf [2]}.
The process $J/\Psi \rightarrow \gamma p \bar{p}$ has also been
considered,
both in the pure quark and in the diquark-quark scheme, in \mbox{{
\bf [3]} ;}
however, in order to avoid ambiguities due to photon radiation by
final quarks,
 occurring in the pure quark case, only the $p \bar{p}$
invariant-mass
distribution, rather than the total decay width, has been computed.
The results derived from either model are found to be of about the
same
magnitude, but neither of them agrees with the data.\\[5pt]

A quark-diquark model of the nucleon has been applied to a
perturbative QCD
description of charmonium decays :
$ \eta_{c},~ \chi_{c0,c1,c2},~ f_{c2} \rightarrow p \bar{p} $ {\bf
[4]}. The
authors obtain a good agreement for the $\chi$'s, but the value of
the
branching ratio for the decay $\eta_{c} \rightarrow p {\bar p}$ is
then found
to be much smaller than the data.\\[5pt]

A different computation of the decay rate of the process
$J/\Psi \rightarrow p {\bar{p}} \gamma $ has been performed { \bf
[5]}.
As in {\bf [3]}, the result is consistently smaller than the data.
\\[5pt]

A different analysis in the timelike region has been performed for
some
exclusive reactions {\bf [6]}, and a new set of parameters has been
obtained.
We shall use it in this paper.\\

Here, introducing a quark-diquark model for baryons in a
perturbative QCD description, we shall discuss the following decays :
$$J/\Psi \rightarrow p  {\bar \Lambda}  K^{-},~
J/\Psi \rightarrow p  {\bar \Sigma^{0}}  K^{-},~
J/\Psi \rightarrow p  {\bar \Sigma(1385)^{0}}  K^{-},$$
$$J/\Psi \rightarrow \Lambda  {\bar \Sigma^{-}}  \pi^{+},~
J/\Psi \rightarrow \Delta(1282)^{++}  {\bar p}  \pi^{-},$$
for which experimental data are available without serious background
problems {\bf [7]}.\\[5pt]

As shown in the typical Feynman diagram of
Fig. 1, the above-mentioned decays of $J/\Psi$ at lowest order
in $ \alpha_s $ are assumed to proceed through the annihilation of
the
$c \bar c$ bound state into three timelike gluons
$g_{1},~g_{2},~g_{3}$ (we use the same names for their respective
four-momenta)
, followed by the materialization of two of them ($g_{1},~g_{3}$)
into two
pairs of quarks, and of the third one ($g_{2}$) into a pair of
diquarks
($D {\bar D}$). Then the produced particles combine so as to form a
baryon, an
antibaryon and a meson in a non-perturbative way. The computation of
this
Feynman diagram is performed by using various assumptions regarding
diquark
form factors and wave functions for baryons and mesons, that will be
presented
hereafter.
\section{The diquark model}
The wave function of the $J/\Psi$ with its four-momentum $k$ (shared
equally
by the constituents $c \; , \; \bar c$), its mass $m_{\psi}$, its
polarisation
four-vector $\varepsilon^{( \lambda )}$ corresponding to the helicity
$\lambda$
($\lambda  = 0, \pm 1 $) and its decay constant $F_{\psi} ( F_{\psi}
\simeq
0.27$ GeV) is given by

$$ \Phi^{(\lambda)}_{\Psi} = \frac{F_\psi}{\sqrt{24}}  ( m_{\psi} +
/\!\!\!k)
\epsilon^{(\lambda)} \frac{1}{\sqrt 3} \sum_{i j } \delta_{i j}$$
where $i , j$ are the colors of $c , \bar c$.\\[5pt]

Omitting color factors and coupling constants, the couplings of a
scalar and a
vector diquark ($D {\bar D}$) pair to a time-like gluon are given by

$$  (S \bar S)_{\mu} =F_s ( D_{\mu^{-}} \bar D_{\mu})$$

$$ (V \bar V)_{\mu} =
F_{1}(\bar D_{\mu}-D_{\mu})\epsilon_{D}^{\sigma*}.\epsilon_{\bar
D}^{\tau}+
F_{2}\{(D.\epsilon_{\bar D}^{\sigma})\epsilon_{D \mu}^{\tau*}-
(\bar{D}.\epsilon_{D}^{\tau*})\epsilon_{\bar{D} \mu}^{\sigma} \}+$$
$$ +F_{3}(\bar D.\epsilon_{D}^{\sigma*})(D.\epsilon_{\bar D}^{\tau})
(D_{\mu} - \bar{D}_{\mu})$$

as written in Ref. {\bf [4]}, where $D_{\mu}$ (${\bar D}_{\mu}$) are
the
diquark (antidiquark) four-momenta,
$\epsilon_{D}^{\sigma}$ ($\epsilon_{\bar D}^{\tau}$) are the diquark
(antidiquark) polarization vectors, $\sigma~(\tau)$ being the
corresponding
helicities.\\[5pt]

We neglect mixed coupling involving both scalar and vector diquarks,
as it is
expected to give only small contributions. \\[5pt]

The diquark form factors are parametrized as in ref.{\bf [6]} for the
gluon\\
\mbox {($g_{2}$)-diquark-antidiquark} vertex (Fig. 1),
\begin{eqnarray*}
F_{S}(g_{2}^{2}) & = &
\left(\frac{Q^{2}_{S}}{Q^{2}_{S} - g^{2}_{2}} \right)
\mbox{~~~~~~~~~},
\mbox{~~~~~}
F_{1}(g_{2}^{2})~=~
\left(\frac{Q^{2}_{V}}{Q^{2}_{V} - g^{2}_{2}} \right)^{2} \\
F_{2}(g_{2}^{2}) & = & (1+k_{v})~F_{1}(g_{2}^{2})
\mbox{~~~~~~},\mbox{~~~~~}
F_{3}(g_{2}^{2})~=~0
\end{eqnarray*}
$k_{v}=1.39$ being the anomalous magnetic moment of the vector
diquark (see
{\bf [6]}). All form factors are restricted to values smaller than
1.3.

As we are in a region of intermediate $g_{i}^{2}$,
the strong-interaction running coupling constant is defined by

$$\alpha_s (g_{i}^{2})=\frac{12 \pi}{25 \ln
(g_{i}^{2}/\Lambda^{2}_{\mbox{QCD}})}$$
with $\Lambda_{\mbox{QCD}} = 200$ MeV, and is restricted to values
smaller
than 0.5.

\section{Wave functions and parameters of the baryons}
We use the parameters $~Q_{0}~\mbox{and}~Q_{1},$
and in addition the normalisation constants $f_{S},~f_{V}$,
determined
through comparison of this model with experimental data. More
precisely, the
set of parameters was obtained by a fit of the proton magnetic form
factor
$G_{M}^{p}$ {\bf [6]} with the following proton distribution
amplitudes (DAs)
which are a kind of harmonic-oscillator wave functions transformed to
the
light cone :

$$\phi_{S}(x) = N_{S}~x~(1-x)^{3}
\exp \biggl[-b^{2} \biggl( \frac{m^{2}_{q}}{x}+
\frac{m^{2}_{S}}{1-x} \biggr) \biggr]$$

$$\phi_{V}(x) = N_{V}~x~(1-x)^{3}~(1+5.8x-12.5x^{2})
\exp \biggl[-b^{2} \biggl( \frac{m^{2}_{q}}{x}
+\frac{m^{2}_{V}}{1-x} \biggr) \biggr]$$

depending on wether the proton is assumed to be made of a quark $q$
and a
scalar diquark $S$, or of a quark $q$ and a vector diquark $V$.
We thus get

$$~f_{S}= 73.85~\mbox{MeV},~Q_{S}^{2}= 3.22~\mbox{GeV}^{2}$$

$$~f_{V}= 127.7~\mbox{MeV},~Q_{V}^{2}= 1.50~\mbox{GeV}^{2}$$

The constituent masses of the u, d quarks and of the diquarks are
taken as :

 $$m_{u}=m_{d}=330~\mbox{MeV} \mbox{~~}, \mbox{~~}
m_{S}=m_{V}=580~\mbox{MeV}$$

The oscillator parameter $b$ is taken to be 0.498 GeV$^{-1}$, and the
constant $N_{S(V)}$ is fixed so that $\int_{0}^{1} \phi(x)~dx=1$.

With this set of parameters the authors of {\bf [6]} fit successfully
the
angular distributions of the protons in the process
$\gamma \gamma \rightarrow  p  {\bar p}$ and the width of
$\eta_{c} \rightarrow p \bar{p}$.

In addition we take for the constituent mass of the s quark
$m_{s}=480~\mbox{MeV}$, while for the masses of scalar and vector
diquarks
made of d and s quarks we take $m_{S(ds)}=m_{V(ds)}=720~\mbox{MeV}$.

\section{Wave functions and parameters of the mesons}
We choose meson DAs which are, as well, harmonic-oscillator wave
function
transformed to the light cone {\bf [9]}.

$$\phi_{\pi(K)}(z) = N_{\pi(K)}~z~(1-z)~
\exp \biggl[-b^{2} \biggl( \frac{m^{2}_{q'}}{z}+
\frac{m^{2}_{ \bar{q}}}{1-z} \biggr) \biggr]$$

where $q'$ and $\bar{q}$ are the quark and antiquark (of
four-momentum $z \pi$
 and $(1-z) \pi$ respectively) forming the meson, and
$N_{\pi(K)}$ is a normalisation factor such that
$\int_{0}^{1} \phi_{\pi (K)}(z)~dz=1$.

As in {\bf [9]} the oscillator parameter $b$ is taken in such a way
that
\mbox{$ \langle k^{2}_{T} \rangle $ = 350 MeV}, $k_{T}$ being the
intrinsic
momentum of either quark inside the meson.

\section{Amplitudes of the final subprocesses}
We evaluate the amplitudes corresponding to different graphs by
factorizing
the various subprocesses, using the covariant indices $\mu, \nu ,
\rho$
for the intermediate timelike gluon four-vectors $g_1, g_2, g_3$.\\

The amplitude of the initial subprocess $ {\cal M}^{\mu \nu \rho} $
for
$J/\Psi$ decay into three gluons is given in Ref. {\bf [2]}. Using
the
definition
$$ J^{\uparrow \uparrow}_{\nu \rho}(q \bar{q}')=
\bar{u}^{\uparrow}(q)
\gamma_{\mu} \gamma_{5} /\!\!\!p \gamma_{\rho} v^{\uparrow}(\bar{q}')
$$
where $q$ and $\bar{q}'$ are the quark and antiquark (of four-momenta
$(1-x)p$ and $(1-y){\bar{p}}$ respectively) contained in the baryon
and
antibaryon, and similar ones for $\mbox{J}^{\uparrow \downarrow}_{\nu
\rho}$,
$\mbox{J}^{\downarrow \uparrow}_{\nu \rho}$,
$\mbox{J}^{\downarrow \downarrow}_{\nu \rho}$,
we shall give the expression of the final amplitudes $ \cal{I}_{\mu
\nu \rho}$
in the ``spin up - spin-up" case (those for the three other cases are
easily
derived therefrom). Those labeled by (a) are exact expressions, while
those
labeled by (b) are approximate ones, neglecting the quark and diquark
masses
in the final subprocesses (which leads to
$J^{\uparrow \downarrow}_{\nu \rho} = J^{\downarrow \uparrow}_{\nu
\rho} = 0$)
and assuming $\phi_{2} = \phi_{3} = \phi_{V}$.

\subsection{$ J/\Psi \rightarrow p \bar{\Lambda}  K^{-} $}

\begin{eqnarray*}
(a)~~~~I^{\uparrow \uparrow}_{\mu \nu \rho} & = &
\frac{f_{S}^{2}}{\sqrt{6}}
\phi_{S}(x) \phi_{S}(y)~
(S(ud)\bar{S}(\bar{u}\bar{d}))_{\mu}
J^{\uparrow \uparrow}_{\nu \rho}(u \bar{s})
\end{eqnarray*}

\begin{eqnarray*}
(b)~~~~I^{\uparrow \uparrow}_{ \mu \nu \rho} & = &
\frac{f_{S}^{2}}{\sqrt{6}}
\phi_{S}(x) \phi_{S}(y)~
(S \bar{S})_{\mu}~J^{\uparrow \uparrow}_{\nu \rho}
\end{eqnarray*}

\subsection{$ J/\Psi \rightarrow p \bar{\Sigma}^{0}  K^{-} $}

\begin{eqnarray*}
(a)~~~~I^{\uparrow \uparrow}_{\mu \nu \rho} & = &
-~\frac{f_{V}^{2}}{9 \sqrt{2}}
\biggl \{ 2~\phi_{2}(x) \phi_{2}(y)~
(V_{+}(ud) \bar{V}_{+}(\bar{u} \bar{d}))_{\mu}
J^{ \downarrow \downarrow }_{\nu \rho}(u \bar{s}) + \\
& + &\phi_{3}(x) \phi_{3}(y)~
(V_{0}(ud) \bar{V}_{0}(\bar{u} \bar{d}))_{\mu}
J^{\uparrow \uparrow}_{\nu \rho}(u \bar{s}) - \\
& - & \sqrt{2} \phi_{2}(x) \phi_{3}(y)~
(V_{+}(ud) \bar{V}_{0}(\bar{u} \bar{d}))_{\mu}
J^{\downarrow \uparrow}_{\nu \rho}(u \bar{s}) - \\
& - & \sqrt{2} \phi_{2}(y) \phi_{3}(x)~
(V_{0}(ud) \bar{V}_{+}(\bar{u} \bar{d}))_{\mu}
J^{\uparrow \downarrow}_{\nu \rho}(u \bar{s}) \biggr \}
\end{eqnarray*}

\begin{eqnarray*}
(b)~~~~I^{\uparrow \uparrow}_{\mu \nu \rho} & = &
-~\frac{f_{V}^{2}}{9
\sqrt{2}}
 \phi_{V}(x) \phi_{V}(y)~\biggl \{2~
(V_{+} \bar{V}_{+})_{\mu}~J^{ \downarrow \downarrow }_{\nu \rho} +
{}~(V_{0} \bar{V}_{0})_{\mu}
J^{\uparrow \uparrow}_{\nu \rho}~\biggr\}
\end{eqnarray*}
\subsection{$J/\Psi \rightarrow \Lambda \bar{\Sigma}^{-}  \pi^{+}$}

\begin{eqnarray*}
(a)~~~~I^{\uparrow \uparrow}_{ \mu \nu \rho} & = &
\frac{1}{6~\sqrt{6}}
\biggl \{ -~3~ f_{S}^{2} \phi_{S}(x) \phi_{S}(y)~
(S(ds)\bar{S}(\bar{d}\bar{s}))_{\mu}
J^{\uparrow \uparrow}_{\nu \rho}( \bar{u} d) + \\
& + & f_{V}^{2} \biggl [ 2~\phi_{2}(x) \phi_{2}(y)
(V_{+}(ds) \bar{V}_{+}(\bar{d} \bar{s}))_{\mu}
J^{ \downarrow \downarrow }_{\nu \rho}( \bar{u} d ) + \\
& + & \phi_{3}(x) \phi_{3}(y)~
(V_{0}(ds) \bar{V}_{0}(\bar{d} \bar{s}))_{\mu}
J^{\uparrow \uparrow}_{\nu \rho}( \bar{u} d ) - \\
& - & \sqrt{2}[\phi_{2}(x) \phi_{3}(y)~
(V_{+}(ds) \bar{V}_{0}(\bar{d} \bar{s}))_{\mu}
J^{\downarrow \uparrow}_{\nu \rho}( \bar{u} d ) + \\
& + & \phi_{2}(y) \phi_{3}(x)~
(V_{0}(ds) \bar{V}_{+}(\bar{d} \bar{s}))_{\mu}
J^{\uparrow \downarrow}_{\nu \rho}( \bar{u} d )] \biggr ] \biggr \}
\end{eqnarray*}

\begin{eqnarray*}
(b)~~~~I^{\uparrow \uparrow}_{ \mu \nu \rho} & = &
\frac{1}{6~\sqrt{6}}
\biggl \{ -~3~ f_{S}^{2}~\phi_{S}(x) \phi_{S}(y)~
(S \bar{S})_{\mu}~J^{\uparrow \uparrow}_{\nu \rho} + \\
& + & f_{V}^{2}~\phi_{V}(x) \phi_{V}(y)\biggl [ 2~
(V_{+} \bar{V}_{+})_{\mu}~J^{ \downarrow \downarrow }_{\nu \rho}
+~(V_{0} \bar{V_{0}})_{\mu}~J^{\uparrow \uparrow}_{\nu \rho}~\biggr ]
\biggr\}
\end{eqnarray*}

\subsection{$ J/\Psi \rightarrow p \bar{\Sigma}(1385)^{0}  K^{-} $}

\begin{eqnarray*}
(a)~~~~I^{\uparrow \frac{3}{2}}_{ \mu \nu \rho} & = &
-~\frac{f_{V}^{2}}{3~
\sqrt{3}} \phi_{2}(x) \phi_{2}(y)~(V_{0}(ud) \bar{V}_{+}(\bar{u}
\bar{d}))_{\mu}
J^{ \uparrow \uparrow }_{\nu \rho}(u \bar{s})
\end{eqnarray*}

\begin{eqnarray*}
(b)~~~~I^{\uparrow \frac{3}{2}}_{ \mu \nu \rho} & = &
-~\frac{f_{V}^{2}}{3~
\sqrt{3}}
\phi_{V}(x) \phi_{V}(y)~(V_{0} \bar{V}_{+})_{\mu}~
J^{ \uparrow \uparrow }_{\nu \rho}
\end{eqnarray*}

\begin{eqnarray*}
(a)~~~~I^{\uparrow \frac{1}{2}}_{ \mu \nu \rho} & = &
\frac{f_{V}^{2}}{9}
\biggl \{-\sqrt{2}~\phi_{2}(x) \phi_{2}(y)~
(V_{+}(ud) \bar{V}_{+}(\bar{u} \bar{d}))_{\mu}
J^{ \downarrow \downarrow }_{\nu \rho}(u \bar{s}) +\\
& + & \sqrt{2}~\phi_{3}(x) \phi_{3}(y)~
(V_{0}(ud) \bar{V}_{0}(\bar{u} \bar{d}))_{\mu}
J^{\uparrow \uparrow}_{\nu \rho}(u \bar{s}) - \\
& - & 2 \phi_{2}(x) \phi_{3}(y)~
(V_{+}(ud) \bar{V}_{0}(\bar{u} \bar{d}))_{\mu}
J^{\downarrow \uparrow}_{\nu \rho}(u \bar{s}) + \\
& + &  \phi_{2}(y) \phi_{3}(x)~
(V_{0}(ud) \bar{V}_{+}(\bar{u} \bar{d}))_{\mu}
J^{\uparrow \downarrow}_{\nu \rho}(u \bar{s}) \biggr \}
\end{eqnarray*}

\begin{eqnarray*}
(b)~~~~I^{\uparrow \frac{1}{2}}_{ \mu \nu \rho} & = &
\frac{f_{V}^{2}}{9}
\phi_{V}(x) \phi_{V}(y)~\sqrt{2}~\biggl \{- ~
(V_{+} \bar{V}_{+})_{\mu}~J^{ \downarrow \downarrow }_{\nu \rho} +
(V_{0} \bar{V}_{0})_{\mu}~J^{\uparrow \uparrow}_{\nu \rho}~\biggr \}
\end{eqnarray*}
\subsection{$J/\Psi \rightarrow \Delta(1232)^{++} \bar{p}  \pi^{-}$}

\begin{eqnarray*}
(a)~~~~I^{\uparrow \frac{3}{2}}_{ \mu \nu \rho} & = &
-~\frac{f_{V}^{2}}{3}
\sqrt{2}~
\phi_{2}(x) \phi_{2}(y)~(V_{0}(uu) \bar{V}_{+}(\bar{u}
\bar{u}))_{\mu}
J^{ \uparrow \uparrow }_{\nu \rho}( u \bar{d} )
\end{eqnarray*}

\begin{eqnarray*}
(b)~~~~I^{\uparrow \frac{3}{2}}_{ \mu \nu \rho} & = &
-~\frac{f_{V}^{2}}{3}
\sqrt{2}~
\phi_{V}(x) \phi_{V}(y)~(V_{0} \bar{V}_{+})_{\mu}
J^{ \uparrow \uparrow }_{\nu \rho}
\end{eqnarray*}

\begin{eqnarray*}
(a)~~~~I^{\uparrow \frac{1}{2}}_{ \mu \nu \rho } & = &
\frac{f_{V}^{2}}{3~
\sqrt{3}} \biggl \{ -~\sqrt{2}~ \phi_{2}(x) \phi_{2}(y)~
(V_{+}(uu) \bar{V}_{+}(\bar{u} \bar{u}))_{\mu}
J^{ \downarrow \downarrow }_{\nu \rho}( u \bar{d} ) +\\
& + & \sqrt{2}~\phi_{3}(x) \phi_{3}(y)~
(V_{0}(uu) \bar{V}_{0}(\bar{u} \bar{u}))_{\mu}
J^{\uparrow \uparrow}_{\nu \rho}( u \bar{d} ) - \\
& - & 2~ \phi_{2}(x) \phi_{3}(y)~
(V_{+}(uu) \bar{V}_{0}(\bar{u} \bar{u}))_{\mu}
J^{\downarrow \uparrow}_{\nu \rho}( u \bar{d} ) + \\
& + &  \phi_{2}(y) \phi_{3}(x)~
(V_{0}(uu) \bar{V}_{+}(\bar{u} \bar{u}))_{\mu}
J^{\uparrow \downarrow}_{\nu \rho}( u \bar{d} ) \biggr \}
\end{eqnarray*}

\begin{eqnarray*}
(b)~~~~I^{\uparrow \frac{1}{2}}_{ \mu \nu \rho
} & = & \frac{f_{V}^{2}}{3~\sqrt{3}}
\phi_{V}(x) \phi_{V}(y)~\sqrt{2}~\biggl \{-~
(V_{+} \bar{V}_{+})_{\mu}~J^{ \downarrow \downarrow }_{\nu \rho} +
{}~(V_{0} \bar{V}_{0})_{\mu}~J^{\uparrow \uparrow}_{\nu \rho}~ \biggr
\}
\end{eqnarray*}

One notes that the coefficients of corresponding terms are the same
here as
in {\bf 5.4}, except for a factor of $\sqrt{3}$.\\[15pt]

Let us remark that, among those reactions, the third one involves
strange
diquarks, i.e. $V(ds)$ and ${\bar{V}}({\bar{d}}{\bar{s}})$. We use
for them
the same parametrization as for the non-strange ones
\section{Partial Widths}

Summing over $\mu,~\nu,~\rho$ and integrating over $x, y, z$, we get
the full
amplitude of the process :
$$ {\cal{M}^{\uparrow \uparrow}}(\mbox{M},\theta) =
\int_{0}^{1} dx \int_{0}^{1} dy \int_{0}^{1} dz~\alpha^{3}_{s}
{\cal{M}}^{\mu \nu \rho}(\mbox{M},\theta,x,y,z)~
I_{\mu \nu \rho}^{\uparrow \uparrow}(\mbox{M},\theta,x,y,z)$$
M being the invariant mass of the baryon-antibaryon system, and
$\theta$
the angular distribution of the baryon in the baryon-antibaryon c.m.
frame,
while we set $\alpha_{s}^{3}=\alpha_{s}(g_{1}^{2})~
\alpha_{s}(g_{2}^{2})~
\alpha_{s}(g_{3}^{2})$.
Using the definition
$${\cal I}^{\uparrow \uparrow}(\mbox{M})= \int_{-1}^{1}
|{\cal M}^{\uparrow \uparrow}(\mbox{M},~\theta)|^{2} d(\cos \theta)$$
and a similar one for ${\cal I}^{\uparrow \downarrow}$, we get the
partial
width :\\
$$\Gamma_{\Psi \rightarrow B_{1} {\bar B}_{2} K^{-}} = \frac{(4 \pi)}
{192 \pi^{3} m_{\Psi}^{3}}
\frac{F_{\Psi}^{2}}{24} \frac{F_{K^{-}}^{2}}{24} C_{F}^{2}
\int_{M_{min}}^{M_{max}} \alpha_{s}^{6}~
p^{*}_{K^{-}} p_{B_{1}} ({\cal I}^{\uparrow \uparrow}
(M)+{\cal I}^{\uparrow \downarrow}(M))~dM $$
with $$ p^{*2}_{K^{-}} = E^{*2}-m^{2}_{K^{-}},~
E^{*2}= \frac{m_{\Psi}^{2}+m_{K^{-}}^{2}-M^{2}}{2 m_{\Psi}},$$
$$p^{2}_{B_{1}}=E^{2}_{B_{1}}-m^{2}_{B_{1}},~
E_{B_{1}}=\frac{M^{2}+m_{B_{1}}^{2}-m_{\bar B_{2}}^{2}}{2 M}$$
and
$$ M_{min}=m_{B_{1}}+m_{\bar B_{2}},~M_{max}=m_{\Psi}-m_{K^{-}}$$ \\
where $m_{B_{1}},~m_{\bar B_{2}}$ are the masses of the baryon and
antibaryon
produced in the decay.\\
Finally $ C_{F} = \displaystyle \frac{5}{2 \times 3^{3}} $ is the
color factor
and $F_{K^{-}}$ the decay constant of the $K^{-}$.
\section{Comparison with experiment}

Let us notice that the decay $J/\Psi \rightarrow p {\bar \Lambda}
K^{-}$
only involves a scalar diquark, while the decays
$J/\Psi \rightarrow p {\bar \Sigma^{0}} K^{-}$,
$J/\Psi \rightarrow p {\bar \Sigma(1385)^{0}} K^{-}$ and
$J/\Psi \rightarrow \Delta(1232)^{++} {\bar p} \pi^{-}$ involve a
vector
diquark exclusively, and
$ J/\Psi \rightarrow \Lambda~\bar{\Sigma}^{-}~\pi^{+} $ involves both
of them.\\

The following table gives the branching ratios R obtained with
formulas (b) of
section 5, to be compared with the experimental data {\bf [10]}. The
values of
R and $\mbox{R}_{exp}$ are multiplied by $10^{3}$.

\vspace*{0.5cm}
\begin{center}
\begin{tabular}{|c|c|c|}
\hline
\mbox{~~~~~~~~~~~~} &  &\\
&  & \\
$ \mbox{J}/\Psi \rightarrow
\mbox{B}_{1}~\bar{\mbox{B}}_{2}~\rm{Meson}$
& ~~ R ~~ & ~~~~ R$_{\mbox{exp}} $ ~~~~ \\[5ex]
\hline
\mbox{~~~~~~~~~~~~} &  &\\
&  &\\
$ J/\Psi \rightarrow \mbox{p}~\bar{\Lambda}~\mbox{K}^{-} $
& 0.08 & 0.89 $\pm$ 0.16\\[1ex]
$ J/\Psi \rightarrow \mbox{p}~\bar{\Sigma}^{0}~\mbox{K}^{-} $
& 0.29 & 0.15 $\pm$ 0.8\\[1ex]
$ J/\Psi \rightarrow \mbox{p}~\bar{\Sigma}^{0}(1385)~\mbox{K}^{-} $
& 0.36 & 0.51 $\pm$ 0.32\\[1ex]
$ J\Psi \rightarrow \Lambda~\bar{\Sigma}^{-}~\pi^{+} $
& 0.10 & 1.06 $\pm$ 0.12\\[1ex]
$ J/\Psi \rightarrow \Delta^{++}(1232)~\bar{\mbox{p}}~\pi^{-} $
& 1.55 & 1.60 $\pm$ 0.50\\[5ex]
\hline
\end{tabular}
\end{center}
\vspace*{0.5cm}
We conclude that, when the scalar diquark is involved (as is the case
in the
first and also predominantly in the fourth decay process here
considered), the
present model does not reproduce the data, i.e. a factor of about 10
is
missing ; notice that this might be due to the fact that we have
taken the
same parameters for strange and non-strange baryons.

On the other hand, in all three cases where the vector diquark is
involved
alone, the agreement is rather satisfactory.

One may hope that in a near future there will be new experimental
results
with higher statistics and thus an improved accuracy. In particular,
it would
be wishable to have experimental data for $d\Gamma/dM$ which would
make it
possible, in a direct way, to extract the form factor of the scalar
resp.
vector diquark from those data ; this would also allow for coherence
tests
between different reactions.

On the other hand, with high statistics, it might also become
possible to
measure angular distributions, i.e. $d^{2}\Gamma/[dM~d(\cos\theta)]$
at fixed
$M$. One should thus be able to check the dynamics of the hard
process, i.e.
the validity of the diquark model as such (either of the
scalar-diquark model
or of the vector-diquark one, depending on the process
considered).\\[15pt]

\noindent{\large \bf{References}}\\[15pt]
{\bf [1]} See e.g. Proceedings of the Workshop on Diquarks, Torino
1988,
eds. M. Anselmino, E. Predazzi (World Scientific Singapore, 1989) ;
M. Szczekovski, Int. J. Mod. Phys. A 4 (1989) 3985  ; M. Anselmino et
al., Lulea preprint TULEA 1992 : 05.\\[10pt]
{\bf [2]} E. H. Kada and J. Parisi, Phys. Rev. D 47 (1993) 3967.
\\[10pt]
{\bf [3]} M. Anselmino, F. Caruso and S. Forte, Phys. Rev. D 44
(1991)
1438.\\[10pt]
{\bf [4]} C. Carimalo and S. Ong, Z. Phys. C 52 (1991) 487.\\[10pt]
{\bf [5]} M. Anselmino and F. Murgia, Z. Phys. C 58 (1993)
429.\\[10pt]
{\bf [6]} P. Kroll, Th. Pilsner, M. Sch|rmann and W. Schweiger, Phys.
Lett. B
316 (1993) 546. \\[10pt]
{\bf [7]} M. W. Eaton et al. (Mark II Collaboration), Phys. Rev. D 29
(1984)
804.\\[10pt]
{\bf [8]} P. Kroll, M. Sch|rmann and W. Schweiger, Int. J. of Mod.
Phys.
A 6 (1991) 4107.\\[10pt]
{\bf [9]} R. Jacob and P. Kroll, Phys. Lett. B 315 (1993)
463.\\[10pt]
{\bf [10]} Particle Data Group, Phys. Rev. D 50 Part I
(1994).\\[15pt]
{\large \bf{Figure caption}}\\[15pt]
Fig 1 : Typical lowest-order diagram for the three-body decay of
$ J/\Psi $ into two baryons and a pseudoscalar meson
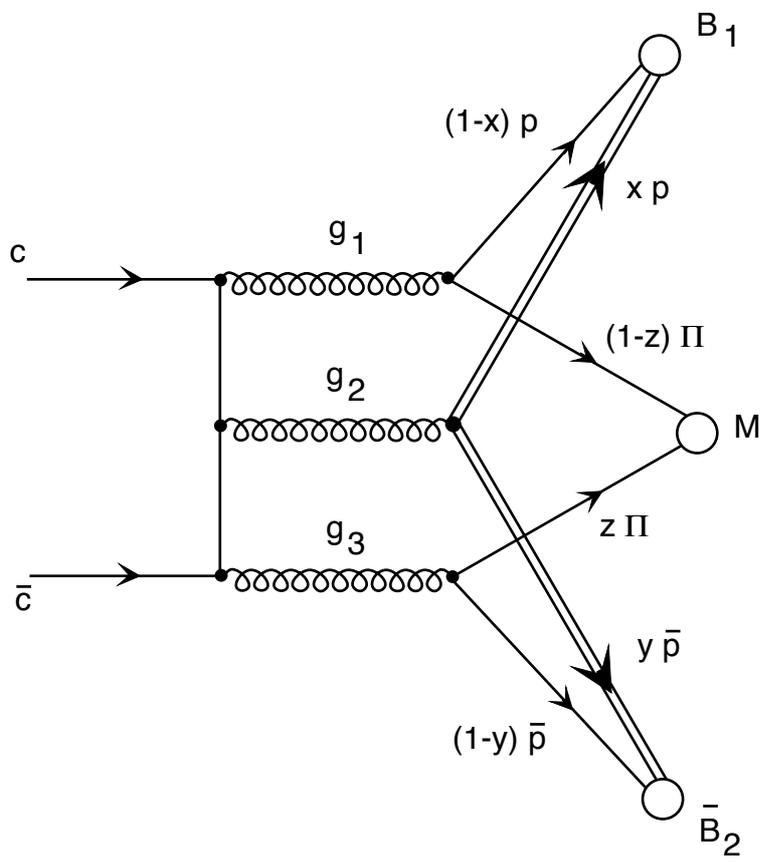
\end{document}